\documentclass[reprint, aps,pra,notitlepage,nofootinbib,longbibliography,floatfix,nobalancelastpage]{revtex4-1}
\usepackage[utf8]{inputenc}
\usepackage{float}
\usepackage{graphicx}  
\usepackage{physics}
\usepackage{esint}
\usepackage{amssymb}  
\usepackage{mathtools}
\usepackage{amsmath}
\usepackage{amsthm}
\usepackage[colorlinks=true,linkcolor=blue,citecolor=magenta,urlcolor=magenta,plainpages=false,pdfpagelabels]{hyperref}
\usepackage{color,xcolor,colortbl}
\usepackage[margin=.75in]{geometry}
\usepackage{bm}
\usepackage[most]{tcolorbox}

\newcommand{\e}{\text{e}}

\newcommand{\cov}[2]{\frac{\text{D}#1}{\text{d}#2}}

\numberwithin{footnote}{section}
\numberwithin{equation}{section}

\begin{document}

%\title{Gravitational effects in two-photon interference \---- Relativistic frame-dragging and the Hong-Ou-Mandel dip}
\title{Relativistic frame-dragging and the Hong-Ou-Mandel dip \---- a primitive to gravitational effects in multi-photon quantum-interference}

\author{Anthony J. Brady}\thanks{Corresponding author: abrady6@lsu.edu}\affiliation{Hearne Institute for Theoretical Physics, Department of Physics and Astronomy, Louisiana State University, Baton Rouge, Louisiana, 70803, USA}

\author{Stav Haldar}\affiliation{Hearne Institute for Theoretical Physics, Department of Physics and Astronomy, Louisiana State University, Baton Rouge, Louisiana, 70803, USA}

\date{\today}

\begin{abstract}
    We investigate the Hong-Ou-Mandel (HOM) effect -- a two-photon quantum-interference effect -- in the space-time of a rotating spherical  
    mass. In particular, we analyze a common-path HOM setup restricted to the surface of the earth and show that, in principle, general-relativistic frame-dragging induces observable shifts in the HOM dip. For completeness and correspondence with current literature, we also analyze the emergence of gravitational time-dilation effects in HOM interference, for a dual-arm configuration. The formalism thus presented establishes a basis for encoding general-relativistic effects into local, multi-photon, quantum-interference experiments. Demonstration of these instances would signify genuine observations of quantum and general relativistic effects, in tandem, and would also extend the domain of validity of general relativity, to the arena of quantized electromagnetic fields. %as a function of the setup's orientation with respect to the rotational axis of the earth.

\end{abstract}

\maketitle
%\tableofcontents

\section{Introduction}\label{sec-intro}

General relativity and quantum mechanics constitute the foundation of modern physics, yet at seemingly disparate scales. On the one hand, general relativity predicts deviations from the Newtonian concepts of absolute space and time -- due to the mass-energy distribution of nearby matter and by way of the equivalence principle \cite{MTW,will2014confrontation} -- which appear observable only at large-distance scales or with high-precision measurement devices \cite{JILA2010}. On the other hand, quantum mechanics predicts deviations from Newtonian concepts of deterministic reality and locality \cite{zeilinger1999rmp} and appears to dominate in regimes in which general relativistic effects are typically and safely ignored. This dichotomous paradigm of modern physics is, however, rapidly changing due to a) the ever-increasing improvement of quantum measurement strategies and precise control of quantum technologies \cite{dowling20032ndquantum, pirandola2018advances} and b) current efforts to extend quantum mechanical demonstrations to large-distance scales -- like, e.g., bringing the quantum to space \cite{pan2019sunhom,pan2019decohere}. 

The maturation of quantum technologies complements the theoretical maturation of quantum fields in curved-space -- a formalism describing the evolution of quantum fields on the (classical) background space-time of general relativity -- which has been finely developed over last fifty-odd years \cite{wald2012history,jacobson2005qfcs}. \emph{Alas, even with such developments, there has not yet been a physical observation requiring the principles of both general relativity and quantum mechanics for its explanation}. With these considerations in mind, the present thus seems a ripe time to furnish potential proof-of-principle experiments capable of and unambiguously exploring the cohesion of general relativity and quantum mechanics, in the near-term. Such is the aim of our work.

In this report, we investigate the emergence of gravitational effects in two-photon quantum interference -- i.e., in Hong-Ou-Mandel (HOM) interference \cite{hom1987OG,ou1988hom,ou2007book} -- for various interferometric configurations. We show, in general, how to encode general-relativistic effects (frame-dragging and gravitational time-dilation) into multi-photon quantum-interference phenomena, and we show, in particular, that the background structure of curved space-time induces observable changes in a HOM-interference signature. %as a function of the geometric orientation of the configurations in question. 

The motivation behind our investigation is twofold. For one, HOM interference is simple to analyze, yet is a primitive to bosonic many-body quantum-interference phenomena \cite{ou2007book,pan2012multiphoton,walschaers2020manybody} and even ``lies at the heart of linear optical quantum computing" \cite{kok2007linear}. Furthermore, there exists a simple yet sharp criterion partitioning the sufficiency of classical and quantum explanations. Namely, if the visibility $\mathcal{V}$ of the HOM-interference experiment is greater than 1/2, a quantum description for the fields is sufficient, whereas a classical description fails to adhere to observations (\cite{ou1988hom,ou2007book}; though, see \cite{sanders2019hom} for recent criticism of the visibility criterion). For two, the emergence of gravitational effects in photonic demonstrations beckons general-relativistic explanations -- whereas, for massive particles, the boundary between general-relativistic and Newtonian explanations is oft-blurred. As rightly pointed-out and emphasized by the authors of refs. \cite{zych2011clock,zych2012photon}, within any quantum demonstration claiming the emergence of general-relativistic effects, a concurrent, clear-cut distinction between classical and quantum explanations \textit{as well as} between Newtonian and general-relativistic explanations must exist, in order to assert that principles of general relativity and quantum mechanics are at play, in tandem. %-- ergo our choice of investigation. 

To place our work in a sharper context, observe that one may broadly decompose the mutual arena of gravitation/relativity (beyond rectilinear motion) and quantum mechanics into four classes: (i) classical Newtonian gravity in quantum mechanics \cite{cow1975gravity, Borner2002quBouncingBall, Asenbaum2017curvWaveFunc,rosi2017Eotvos}, (ii) non-inertial reference frames in quantum mechanics \cite{cow1980sag, Bonse1983aCOW,Fuentes2017NIRexp, padgett2019hom, padgett2020nir}, (iii) classical general relativity in quantum mechanics  \cite{zych2011clock,zych2012photon,rideout2012,brodutch2015,zych2018QEEP,terno2018proposal,roura2020PRX,Tapia2020BellGravity}, and (iv) the quantum nature of gravity \cite{vedral2017QGwitness,bose2017QGwitness,serafini2018omgravity,krisnanda2020observable,RovelliVedral2020}. Although there has been a great deal of theoretical investigation in these areas, there has only been observational evidence for (i) and (ii) (see, e.g., refs. \cite{cow1975gravity, Borner2002quBouncingBall, Asenbaum2017curvWaveFunc,rosi2017Eotvos} and \cite{cow1980sag,Bonse1983aCOW,Fuentes2017NIRexp, padgett2019hom}, respectively), which indeed has supported consistency between these formalisms when they are concurrently considered -- e.g., consistency between Newton's theory of gravitation and standard non-relativistic quantum theory, in regimes where both are prevalent. Our work serves as a contribution to (iii) -- where one must describe the gravitational field by a classical metric theory of gravity and physical probe-systems via quantum mechanical principles.

We compartmentalize our paper in the following way. In Section \ref{sec-qu_optics}, we provide a model for the electromagnetic field in curved space-time, in the weak-field regime and proceed to quantize the field under suitable approximations. One can consider this formalism, in succinct terms, as \emph{geometric quantum-optics in curved space-time, in the weak-field regime}. In Section \ref{subsec-metric}, we discuss the metric local to an observer and its relation to the background Post-Newtonian (PN) metric, thus prescribing the space-time upon which the quantum field of section \ref{sec-qu_optics} propagates. In Section \ref{subsec-hom}, we provide some background on HOM interference and comment on interferometric noise and gravitational decoherence in context. In Section \ref{subsec-setup_results}, we combine and apply the formalism of previous sections to various interferometric configurations, explicitly showing that gravitational effects, in principle, induce observable changes in HOM-interference signatures. We also provide order-of-magnitude estimates for these effects and briefly discuss observational potentiality. Finally, in Section \ref{sec-conclusion}, we summarize our work.

\begin{figure*}[t]
    \centering
    \includegraphics[width=\textwidth]{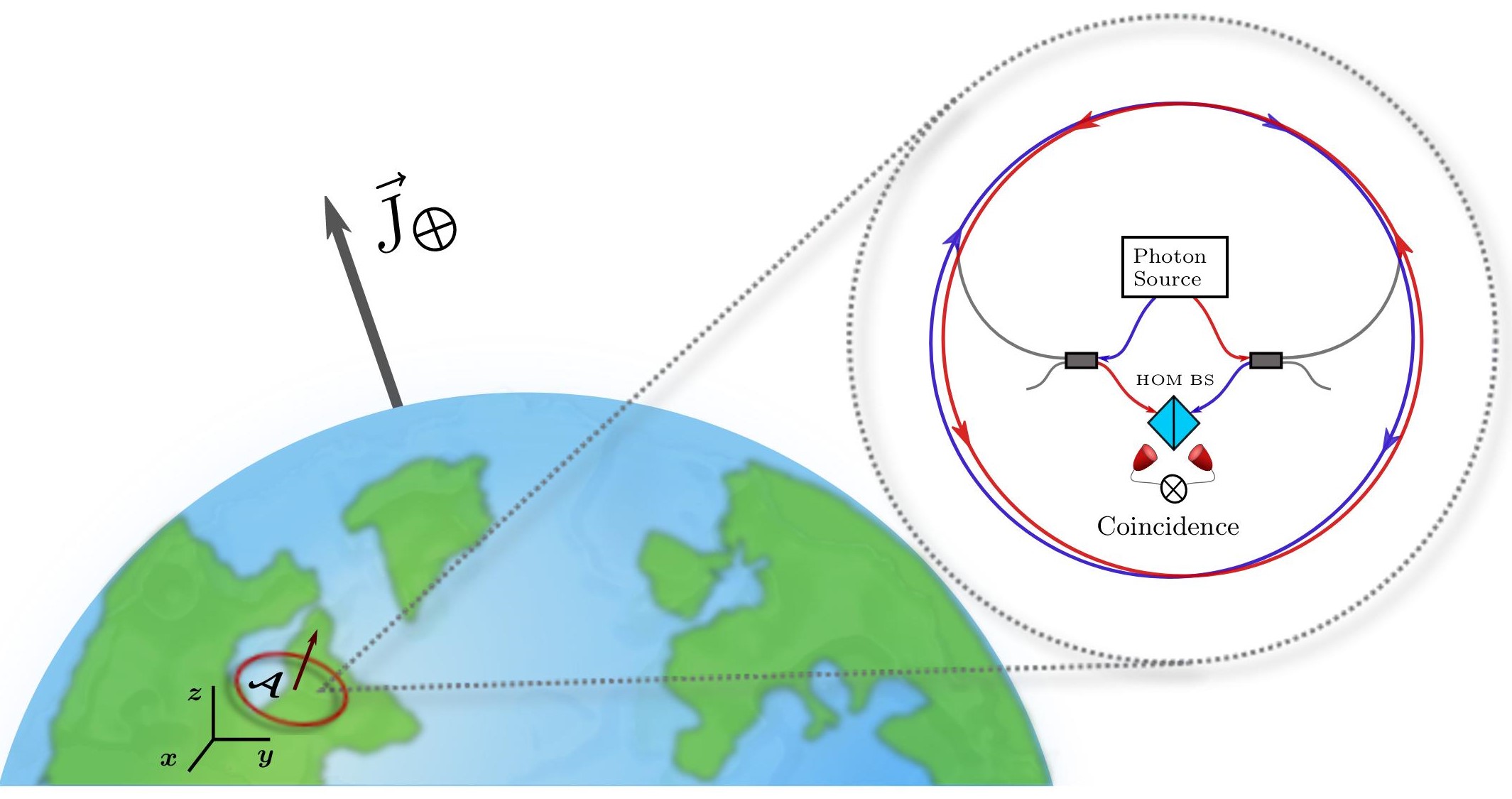}
    \caption{Schematic of a Hong-Ou-Mandel interferometer, in a common-path configuration, located on the earth's surface. The gray boxes just after the photon source are fiber-coupler and beamsplitter systems, with high transmissivity, used to ensure common-path propagation (cf. \cite{padgett2019hom, padgett2020nir}).}
    \label{fig-earth_hom}
\end{figure*}

\section{Methods}\label{sec-methods}

\subsection{Modeling quantum optics in curved space-time}\label{sec-qu_optics}
\subsubsection{Geometric optics in curved space: a brief review}\label{subsubsec-geo_optics}
We model the electromagnetic field by a free, massless, scalar field $\phi$ satisfying the (minimally coupled) Klein-Gordon equation

\begin{equation}
    \nabla^\mu\nabla_\mu\phi=0\label{eq-kg_eq},
\end{equation}
where $\nabla_\mu$ is the metric compatible covariant derivative of general relativity. Such a model disregards polarization-dependent effects but accurately accounts for the amplitude and phase of the field.

We assume the field to be quasi-monochromatic in the geometric optics approximation \cite{born1999optics,MTW} so that it can be rewritten (in complex form) as
\begin{equation}
\phi_k=\alpha_k^\prime\text{e}^{\text{i}S^\prime_k} , \label{eq-field}  
\end{equation}
with $S^\prime$ the eikonal (or phase), $\alpha^\prime$ a slowly-varying amplitude with respect to variations of the eikonal, and $k$ denoting the wave-vector mode which $\phi_k$ is centered about. The wave equation \eqref{eq-kg_eq} in the geometric optics approximation yields transport equations along the vector field $\nabla_\mu S^\prime_k$,

\begin{align}
    \left(\nabla^\mu S^\prime_k\right)\left(\nabla_\mu S^\prime_k\right)&=0 \label{eq-transport1}\\
    \left(\nabla^\mu S^\prime_k\right)\left(\nabla_\mu \ln{\alpha^\prime_k}\right)&=-\frac{1}{2}\nabla^\mu\nabla_\mu S^\prime. \label{eq-transport2}
\end{align}
The first equation implies that $\nabla_\mu S^\prime_k$ is a null vector. One can show that it also satisfies a geodesic equation \[\left(\nabla^\mu S^\prime_k\right)\nabla_\mu\left(\nabla_\nu S^\prime_k\right) =0. \] Thus, the transport equations, \eqref{eq-transport1} and \eqref{eq-transport2}, determine how the eikonal and amplitude, respectively, evolve along the null ray drawn out by $\nabla_\mu S^\prime_k$. Observe that the evolution of the eikonal along the ray is sufficient to determine the field $\phi$ entirely. These are the main results of geometric optics. 

Lastly, from the transport equations, one may derive a conservation equation 
\begin{equation}
    \nabla^\mu\left(\nabla_\mu S^\prime_k\abs{\alpha^\prime_k}^2\right)=0,
\end{equation}
which, via Gauss's theorem, leads to a conserved `charge'
\begin{equation}
    Q_k\coloneqq\int_\Sigma\dd\Sigma^\mu \left(\nabla_\mu S^\prime_k\right)\abs{\alpha^\prime_k}^2, \label{eq-photon_conserv}
\end{equation}
with $\Sigma$ being a spacelike hypersurface. Physically, the conservation equation is a conservation of photon flux such that the number of photons $Q_k$ is a constant for all time. 

The equations thus posed are precisely that of geometric optics in curved space-time, excepting a transport equation for the polarization vector of the field (see, e.g., refs. \cite{born1999optics,MTW}). 

\subsubsection{Geometric optics in the weak-field regime}
We work with laboratory or near-earth experiments in mind. Therefore, it is sufficient to consider general relativity in the weak-field regime.  

We assume local gravitational fields, accelerations, etc. are sufficiently weak and consider first-order perturbations $h_{\mu\nu}$ about the Minkowski metric $\eta_{\mu\nu}$ such that the space-time metric $g_{\mu\nu}$, in some local region in space, can be written as
\begin{equation}
    g_{\mu\nu}=\eta_{\mu\nu}+h_{\mu\nu}.
\end{equation}
The metric perturbation induce perturbations of the eikonal and amplitude about the Minkowski values $(S_k,\alpha_k)$, i.e.
\begin{align}
    S^\prime_k&=S_k+\delta S_k\\
    \alpha^\prime_k&=\alpha_k(1+\varepsilon_k),
\end{align}
with $(\delta S_k,\varepsilon_k)$ being the leading perturbations of $\order{h}$. Defining $k_\mu\coloneqq\partial_\mu S_k$ as the Minkowski wave-vector, we show that the general transport equations, \eqref{eq-transport1} and \eqref{eq-transport2}, imply the $\order{1}$ transport equations
\begin{align}
    k^\mu k_\mu=0 &\implies k^\mu\partial_\mu k_\nu=0 \label{eq-kmu}\\
    k^\mu\partial_\mu\alpha_k&=-\frac{1}{2}\partial_\mu k^\mu \label{eq-amplitude_mink}  ,
\end{align}
which are the geometric optics equations in Minkowski space-time , and $\order{h}$ transport equations
\begin{align}
    k^\mu\partial_\mu\delta S_k&=-\frac{1}{2}k^\mu k^\nu h_{\mu\nu} \label{eq-deltaS_evolve} \\
    k^\mu\partial_\mu\varepsilon_k &= -\frac{1}{2}\left(\partial_\mu\partial^\mu\delta S_k+ \Gamma^\mu_{\mu\nu}k^\nu\right)\label{eq-epsilon_evolve},
\end{align}
with \[{\Gamma^\mu}_{\beta\nu}=\frac{1}{2}\eta^{\mu\kappa}\left(\partial_\beta h_{\nu\kappa} +\partial_\nu h_{\beta\kappa} -\partial_\kappa h_{\beta\nu} \right),\] being the torsion-free connection coefficients of general relativity to $\order{h}$ \cite{MTW}. Therefore, one observes that the null ray $k^\mu$, together with the metric perturbation $h_{\mu\nu}$, uniquely determines the eikonal $S_k^\prime$, the amplitude $\alpha_k^\prime$ [per eqs. \eqref{eq-kmu}-\eqref{eq-epsilon_evolve}], and hence, the field $\phi$ [eq. \eqref{eq-field}]. 

\paragraph{Generic solutions:} Let us suppose that the null curve $\gamma$ is parameterized by parameter $\lambda$ and has (Minkowski) coordinate expression $x^\mu(\lambda)$ such that $k^\mu=\dv*{x^\mu}{\lambda}$. This implies $\dd\lambda=\dd x^0/k=\dd\Vec{x}\vdot\Vec{k}/k^2
$, with $k^2\coloneqq\Vec{k}\vdot\Vec{k}$ the Euclidean scalar product. With this supposition, we find general solutions to the perturbative transport equations
\begin{align}
    \delta S_k&=-\frac{1}{2}\int_{\gamma}\dd x^\mu k^\nu h_{\mu\nu}\label{eq-phase_soltn} \\
    \varepsilon_k&=-\frac{1}{2}\int_{\gamma}\dd\lambda \left(\partial_\mu\partial^\mu\delta S_k+ {\Gamma^\mu}_{\mu\nu}k^\nu\right).
\end{align}
Observe that, to first non-trivial order, 
\begin{equation*}
    (1+\varepsilon_k)=\exp[-\frac{1}{2}\int_{\gamma}\dd\lambda \left(\partial_\mu\partial^\mu\delta S_k+ {\Gamma^\mu}_{\mu\nu}k^\nu\right)].
\end{equation*}
Combining this with the $\order{1}$ amplitude and using the fact that \[ \nabla_\mu\nabla^\mu S_k^\prime=\partial_\mu k^\mu + \partial_\mu\partial^\mu\delta S_k+ {\Gamma^\mu}_{\mu\nu}k^\nu,\] to $\order{h}$, we obtain
\begin{equation}
    \alpha_k^\prime=\exp[-\frac{1}{2}\int_\gamma\dd\lambda\left(\nabla_\mu\nabla^\mu S_k^\prime\right)],\label{eq-amplitude_soltn}
\end{equation}
where the initial amplitude is taken as unity, for simplicity. This is analogous to the generic solution for the amplitude in curved space-time -- excepting that the geometric curve, over which the integral is evaluated, is the Minkowski null geodesic and not the general null geodesic generated by $\nabla_\mu S_k^\prime$. 

\subsubsection{Approximately orthonormal modes and quantization}

We wish to investigate local (e.g. in a finite region of space) quantum optics experiments on a curved background space-time, in the weak-field regime. Hence, we must produce a quantum description for the field. To do so, we introduce a set of approximately orthonormal solutions to the Klein-Gordon equation \eqref{eq-kg_eq}, which -- upon quantization -- allows us to define creation and annihilation operators of the quantum field obeying the usual commutation relations \cite{jacobson2005qfcs}. From this construction, a Fock space can be built and quantum optics experiments analyzed per usual. We note that the following formalism describes `local' field solutions which are valid descriptions of the field, in some finite region of space (of size $\ll \ell$, see below), to some specified accuracy [to $\order{h^2}$, see below]. 
\paragraph{Approximately orthonormal modes:}
From the generic solutions, define the mode $u_k$ which satisfies the Klein-Gordon equation in the weak-field regime
\begin{equation}
    u_k\coloneqq\mathcal{N}_k\alpha_k^\prime\text{e}^{\text{i}(k^\mu x_\mu +\delta S_k)},
\end{equation}
where $\mathcal{N}_k$ is a normalization constant. We prove that, to some error, the set $\{u_k\}$ define an orthonormal set of classical solutions to the Klein-Gordon equation, with respect to the Klein-Gordon inner product \cite{jacobson2005qfcs}
\begin{equation}
    \left(f,g\right)_{KG}\coloneqq\text{i}\int_\Sigma\dd\Sigma^\mu \left(f^*\partial_\mu g-g\partial_\mu f^*\right),
\end{equation}
where $\Sigma$ ia a suitable normalization volume, i.e. a spatial hypersurface of sufficient size. We assume plane-waves for the Minkowski modes. The sketch of the proof goes as follows. 

We wish to show that 
\begin{equation}
    \left(u_k,u_{k^\prime}\right)_{KG}\approx \delta(\Vec{k}-\Vec{k}^\prime),
\end{equation}
to $\order{h^2}$. Consider $h_{\mu\nu}$ to be time-independent and to vary over a characteristic length scale $\partial h\sim\ell^{-1}$, where, for example, $\ell\sim c^2/g\sim1\text{ly}$ for a terrestrial gravitational acceleration $g\approx9.8 \ \text{m}/\text{s}^{2}$. We further assume the integration volume $\Sigma$ to obey $\Sigma^{1/3}\ll\ell$, i.e. $h\ll1$ within the spatial region of interest (the laboratory). Then, analyzing the Klein-Gordon inner product perturbatively, we see that 
\begin{equation}
    \left(u_k,u_{k^\prime}\right)_{KG}\approx\left(u_k,u_{k^\prime}\right)_{KG}^{(0)}+\left(u_k,u_{k^\prime}\right)_{KG}^{(1)},
\end{equation}
where $\left(u_k,u_{k^\prime}\right)_{KG}^{(0)}\propto\delta(\Vec{k}-\Vec{k}^\prime)$ is the flat-space inner product for the plane-wave modes\footnote{We approximate a continuum for the normalization volume, whilst maintaining the constraint $\Sigma^{1/3}\ll\ell$.} and $\left(u_k,u_{k^\prime}\right)_{KG}^{(1)}$ is the inner product containing terms to $\order{h}$, which can be written as 

\begin{equation}
    \left(u_k,u_{k^\prime}\right)_{KG}^{(1)}\propto\text{e}^{-\text{i}(\omega_k-\omega_{k^\prime})t}\int_\Sigma\dd^3\Vec{x} \ \text{e}^{\text{i}(\Vec{k}-\Vec{k}^\prime)\vdot\Vec{x}}f(h), \label{eq-order1_ip}
\end{equation}
with $(t,\Vec{x})$ being Lorentz coordinates on $\Sigma$ and $f(h)$ a function of $\order{h}$ that varies over the length scale $\ell$. It is then sufficient to show that $\left(u_k,u_{k^\prime}\right)_{KG}^{(1)}$ only has support at $k=k^\prime$. This is done in two steps. For $k-k^\prime\gg\ell^{-1}$, the phase rapidly oscillates with respect to $f(h)$ over the entire integration volume. Thus, by the Reimann-Lebesgue lemma, the integral averages to zero. On the other side, for $k-k^\prime\sim\ell^{-1}$, we have $(k-k^\prime)x\sim\order{h} \ \forall \ x\in\Sigma$. Thus, we may set $k=k^\prime$ in the integral, to accuracy $\order{h^2}$. Therefore, the integral \eqref{eq-order1_ip} only has support at $k=k^\prime$ to $\order{h^2}$, QEI.  

\paragraph{Canonical quantization:}We decompose the field in the basis set $\{u_k\}$,

\begin{equation}
    \phi = \int\dd^3\vec{k}\left(a_ku_k +\text{c.c}\right),
\end{equation}
with basis coefficients, $a_k$, found from the Klein-Gordon inner product
\begin{align}
    a_k & \approx (u_k,\phi)_{KG} \\
    a_k^* & \approx (\phi,u_k)_{KG}.
\end{align}
We proceed to canonically quantize the field by introducing the canonical momentum for the classical Klein-Gordon field \cite{jacobson2005qfcs}
\begin{equation*}
    \pi=\abs{\mathfrak{h}}^\frac{1}{2}\partial_\tau\phi,
\end{equation*}
which is defined on a future-oriented spatial hypersurface $\Sigma_\tau$, with $\abs{\mathfrak{h}}$ the determinant of the induced spatial metric on $\Sigma_\tau$. Note that $\Sigma_\tau$ is simply the surface of simultaneity at time $\tau$. 

The field $\phi$ and canonical momentum $\pi$ are then promoted to Hermitian operators, $\phi\rightarrow\hat{\phi}$ and $\pi\rightarrow\hat{\pi}$, set to satisfy the equal-time canonical commutation relation 
\begin{equation}
    \commutator{\hat{\phi}(\vec{x})}{\hat{\pi}(\vec{y})}\coloneqq  i\delta(\vec{x}-\vec{y}),\label{eq-canonical_comm}
\end{equation}
per the correspondence principle. Here, $(\vec{x},\vec{y})$ are spatial coordinates on $\Sigma_\tau$. Under this prescription, the field operator has a basis decomposition 
\begin{equation}
    \hat{\phi}=\int\dd^3\vec{k}\left(\hat{a}_ku_k +\hat{a}^\dagger_ku_k^*\right),
\end{equation}
with $(\hat{a},\hat{a}^\dagger)$ the annihilation and creation operators, which are found per above
\begin{align}
     \hat{a}_k &\approx (u_k,\hat{\phi})_{KG} \\
    \hat{a}_k^\dagger &\approx (\hat{\phi},u_k)_{KG}.
\end{align}
Given the basis decomposition and canonical commutation relation \eqref{eq-canonical_comm}, one can show that 
\begin{align}
    \commutator{\hat{a}_k}{\hat{a}_q^\dagger}&=(u_k,u_q)_{KG}\\
    &\approx\delta(\vec{k}-\vec{q}).
\end{align}
\paragraph{Fock space and wave packets:} The (approximate) orthonormality of the classical mode solutions, together with the commutation relations, is sufficient to build a Fock space spanned by states of the form \cite{jacobson2005qfcs} 
\begin{equation}
    \bigotimes_k\frac{(\hat{a}_k^\dagger)^{n_k}}{\sqrt{n_k!}}\ket{\mathbf{0}},
\end{equation}
with \[\ket{\mathbf{0}}\coloneqq\bigotimes_k\ket{0_k},\] being the vacuum state on $\Sigma_\tau$. Thus, the usual interpretation of the creation operator $\hat{a}_k^\dagger$ follows -- e.g., as creating a single-photon in the mode $u_k$. Note, however, that the Fock states above are non-normalizable (or normalizable up to a Dirac-delta distribution). We remedy this difficulty by constructing wave-packets. As example, we construct a positive frequency wave-packet,
\begin{equation}
    f=\int\dd^3\vec{k} \ f_k^*u_k,
\end{equation}
with $f_k\in\mathbb{C}$, usually peaked around some wave-vector, but more suitably an $L^2$ function in $k$-space. From thus, we define the annihilation wave-packet operator associated with $f$,
\begin{equation}
    \hat{a}_f\coloneqq(f,\hat{\phi})_{KG}\approx\int\dd^3\vec{k} \ f_k\hat{a}_k.
\end{equation}
Commutation relations for the wave-packet operators then follow
\begin{equation}
\begin{split}
    \commutator{\hat{a}_f}{\hat{a}^\dagger_f}&=(f,f)_{KG} \\&\approx\int\dd^3\vec{k} \ \abs{f_k}^2=1,
\end{split}
\end{equation}
where approximate orthonormality of the mode functions was assumed and the $L^2$ property of $f_k$ was used. Per above, one can build a Fock space of wave-packet states which are properly normalizable. Furthermore, this construction permits the interpretation of $\hat{a}^\dagger_f$ as creating a photon occupying the wave-packet $f$.

\subsection{The Metric}\label{subsec-metric}
In order to analyze gravitational effects in quantum optics experiments in a laboratory environment, we must prescribe the metric local to a laboratory observer, which can be done in the so-called proper reference frame -- a reference frame naturally adapted to an arbitrary observer in curved space-time. We follow references \cite{MTW, delva2017universe} in the construction of the proper reference frame and its relation to the background PN frame, by providing key equations and brief explanations, whilst leaving explicit details to the references therein.

\subsubsection{Proper reference frame: the local metric}\label{subsubsec-prop_ref}

\begin{figure}[t!]
    \centering
    \includegraphics[width=\linewidth]{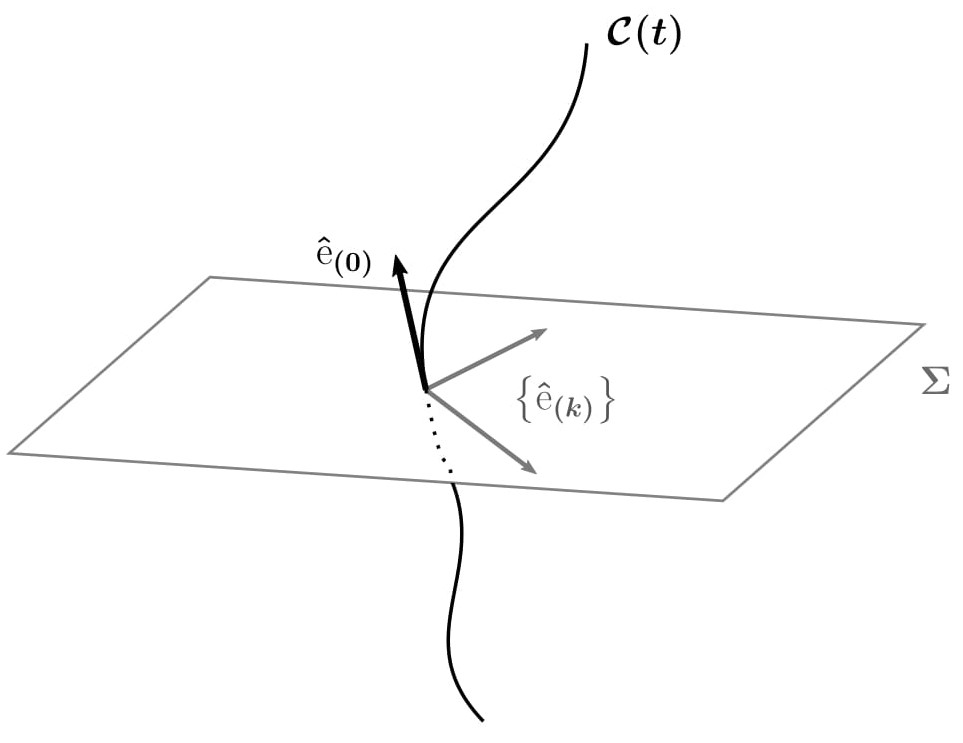}
    \caption{Schematic of the observer's world-line $\mathcal{C}$ through space-time, with $\hat{\e}_{(0)}$ parallel to the observer's four-velocity and $\{\hat{\e}_{(k)}\}$ designating the spatial triad attached to the observer. All measurements are made, by the observer, with respect to these basis vectors.}
    \label{fig-obs_worldline}
\end{figure}

Consider an ideal observer moving along their world-line $\mathcal{C}(t)$, where $t$ is the (proper) time the observer measures with a good clock. The observer also `carries' with them a space-time tetrad $\{\hat{e}_{(\mu)}\}$, associated with coordinates $x^{(\mu)}$, such that: i) the time coordinate satisfies $x^{(0)}=t$, which implies $\hat{e}_{(0)}\coloneqq\partial_{(0)}t$ is the observer's four-velocity (tangent to $\mathcal{C}$), and ii) the set $\{\hat{e}_{(k)}\}$ is a rigid set of spatial axes `attached' to the observer (FIG. \ref{fig-obs_worldline}). Thus, the observer measures the passage of time, spatial displacements, angles, local accelerations, rotations, etc. with respect to these basis vectors. The tetrad is chosen in such a manner that: i) the local metric, $g_{(\mu)(\nu)}$, reduces to the Minkowski metric on $\mathcal{C}$,\footnote{A representation of the notion that space-time is locally flat \cite{MTW}.} \[g_{(\mu)(\nu)}\big|_\mathcal{C}=\eta_{(\mu)(\nu)},\] and ii) the metric, in a neighborhood of $\mathcal{C}$, describes that of an accelerating and rotating reference frame in flat space-time, which, to linear displacements in $x^{(k)}$, takes the form
\begin{equation}
    \dd s^2= -\Big(1+2\vec{\gamma}\vdot\vec{x}\Big)\dd t^2 + 2\left(\vec{\omega}^\prime\cross\vec{x}\right)\vdot\dd\vec{x}\dd t + \dd\vec{x}^2,\label{eq-prop_metric}
\end{equation}
where $(\vec{\gamma},\vec{\omega}^\prime)$ are the acceleration and angular velocity that the observer measures with, e.g., accelerometers and gyroscopes. Note that, these accelerations and angular velocities are independent of spatial coordinates $x^{(k)}$; though, in general, they may depend on the observer's proper time, $t$. 

At this juncture, some general observations and comments about the proper reference frame should be made. First, one should apparently associate the (small) linear displacements from the observer's world-line, as considered in this section, with the ``weak field regime", considered in previous parts of the paper. Second, in the context of general relativity, accelerations and rotations are coordinate-dependent quantities. Therefore, for a free-falling (parallel-transported) observer, such quantities vanish entirely. Then, the first non-trivial components in the metric are at $\order{x^2}$ and depend on Reimann curvature \cite{delva2017universe} (see also Ch. 13 of \cite{MTW}). All this is not to say that the above construction is invalid or not useful. On the contrary, the proper reference frame is extensive enough to accommodate for local gravitational time-dilation and general-relativistic frame-dragging effects. Thirdly and lastly, the proper reference frame is quite generic and does not, necessarily, relate to gravitational effects. It is only once Einstein's field equations are considered and a background space-time prescribed therefrom that the local dynamics, observed about $\mathcal{C}$, can be attributed to gravitational phenomena. %attributed generically to accelerations and rotations. 

\subsubsection{Relation to background PN metric}\label{subsubsec-ppn_metric}
One may use the PN formalism (see \cite{will2014confrontation} and Ch. 39 of \cite{MTW}) to describe the space-time metric near massive bodies in the weak field, slowly moving, and slowly evolving regime -- which is sufficient to analyze, e.g., solar-system experiments aiming to test metric theories of gravity. From the viewpoint of general relativity, the idea is to expand, in successive orders of a small parameter $\epsilon$, and solve Einstein's field equations, thereby obtaining a sufficiently accurate background space-time metric. The small parameter $\epsilon$ is set by physical quantities -- the Newtonian potential $U$ and the velocity $v$ of the body in the PN frame -- such that \[U, \ v^2\sim \order{\epsilon^2}.\] The condition of ``slowly evolving", for all quantities $\mathcal{Q}$, is set by $\partial_t\mathcal{Q}/\partial_i\mathcal{Q}\sim\order{\epsilon}$. 

Now, for example, letting bare subscripts indicate tensor components in the PN frame, the space-time metric for a solid rotating sphere (the earth) in the PN formalism is   
\begin{equation}
g_{\mu\nu}:
\begin{cases} 
      g_{00}=-\left(1-2U\right) +\order{\epsilon^4} \\
      
      g_{0i}=-4V_i + \order{\epsilon^5}\\ 
      g_{ij}=\left(1+2U\right)\delta_{ij} +\order{\epsilon^4},
   \end{cases}\label{eq-pn_metric}
\end{equation}
where
\begin{align}
    U&= \frac{GM}{r} \\
     \vec{V}&=\frac{\vec{J}\cross\vec{r}}{2r^3},
\end{align}
and $\vec{J}=I\vec{\omega}$, with $I$ being the moment of inertia and $\vec{\omega}$ the rotation rate of the earth in the PN frame. It follows that the off diagonal metric components are of $\order{\epsilon^3}$. From here, one can relate the PN space-time metric to physical quantities that an (proper) observer locally measures, i.e. to the accelerations and rotations $(\vec{\gamma},\vec{\omega}^\prime)$. One accomplishes this in the following manner (see \cite{delva2017universe} for explicit details).  

Let parenthetic subscripts indicate tensor components in the proper reference frame of an observer constrained to the surface of the earth. Thus, the spatial triad $\{\hat{\e}_{(k)} \}$ physically correspond to rigid Cartesian axes stuck to the earth's surface -- where, e.g., two basis vectors may point along increasing longitudinal and latitudinal lines at the observer's position and the third along the radial normal. Then, the following transport equation governs the evolution of the observer's tetrad along their world-line,
\begin{equation}
    \cov{\hat{\e}_{(\alpha)}^\mu}{t} =- \hat{\e}_{(\beta)}^\mu{\Omega^{(\beta)}}_{(\alpha)},
\end{equation}
where $\text{D}/\text{d}t$ is the covariant derivative along the observers world-line and $\hat{\e}_{(\alpha)}^\mu$ are the vector components of the observer's tetrad basis in the PN frame. The anti-symmetric transport tensor $\Omega_{(\alpha)(\beta)}$ encodes gravitational and non-inertial effects and is found via
\begin{equation}
    \Omega_{(\alpha)(\beta)}=\eval{\frac{1}{2}g_{\mu\nu} \left(\hat{\e}^\mu_{(\alpha)}\cov{\hat{\e}^\nu_{(\beta)}}{t} -\hat{\e}^\nu_{(\beta)}\cov{\hat{\e}^\mu_{(\alpha)}}{t} \right)}_\mathcal{C}.
\end{equation}
The components of the transport tensor correspond precisely to the locally measured accelerations and rotations, i.e.
\begin{align*}
\Omega_{(0)(j)}&\coloneqq\gamma_{(j)}\\
   \Omega_{(i)(j)}&\coloneqq\varepsilon_{(i)(j)(k)}\omega^\prime_{(k)},
\end{align*}
with $\varepsilon_{ijk}$ being the Levi-Civita symbol. From thus, one finds \cite{delva2017universe}
\begin{equation}
    \vec{\gamma}=\vec{\mathfrak{a}}-\grad{U},
\end{equation}
with $\vec{\mathfrak{a}}$ being the coordinate (e.g., centrifugal) acceleration. Note that, for an earthbound observer $\grad{U}=\vec{g}$, where $g\approx9.8\text{m}/\text{s}^2$ is the gravitational acceleration on earth's surface. 

One also finds an explicit expression for the rotation vector, 
\begin{widetext}
\begin{equation}
    \vec{\omega}^\prime = \vec{\omega}\left(1+\frac{1}{2}v^2+U\right) + \frac{1}{2}\vec{v}\cross\vec{\gamma} - \left(\frac{3}{2}\vec{v}\cross\grad{U}+2\grad\cross\vec{V}\right).\label{eq-rotation_rate}
\end{equation}
\end{widetext}
The first term is the rotation rate of the earth, as measured by an earthbound observer. The second term is the Thomas precession term. Finally, the third and fourth terms are the geodetic and Lense-Thirring terms, respectively. Considering the coordinate velocity (and acceleration, $\mathfrak{a}$) as being due solely to the rotational motion of the earth (hence, $\vec{v}=\vec{\omega}\cross\vec{R}$ with $R$ being the earth's radius) one finds the centrifugal component of the Thomas precession term, $\vec{v}\cross\vec{\mathfrak{a}}$, to be 1000 times smaller than the gravitational terms. We thus ignore this term hereafter. For brevity, we shall also let $\vec{\omega}$ denote the rotational rate as measured by an earthbound observer. 

\subsection{Two-photon interference and the Hong-Ou-Mandel dip}\label{subsec-hom}

The HOM effect is a two-photon quantum-interference effect, which embodies the general phenomena of boson (photon) bunching -- the inclination for identical bosons to congregate -- as a consequence of the commutation relations for a bosonic field, and can be used to quantify the distinguishability of  single photons \cite{pan2012multiphoton}. For example, a primitive HOM configuration consists of a two-photon source, which creates independent single-photon wavepackets, and a 50/50 beamsplitter, where the photons interfere. At the output ports of the beamsplitter, one positions photodetectors, records coincident detection events, and quantifies the probability of a coincident detection -- i.e., the likelihood that \emph{both} detectors register a single-photon event. For identical single-photons -- identical polarization, spectral and temporal profiles etc. -- the probability of a coincident detection is zero, due to the tendency of the photons to bunch in one output port of the beamsplitter or the other. If, however, we consider spectral wavepackets, we can induce distinguishability by introducing a relative time delay between the wavepackets prior to the beamsplitter, \textit{ceteris paribus}. As a consequence, the likelihood of a coincident detection event increases, as the relative time delay departs from zero. The transition of the coincidence probability, from zero to a nonzero value, leads to a dip-like structure in its functional behavior, with respect to the time delay (see, e.g., inset of FIG. \ref{fig-sagnac_results}). This dip in coincidence events is the well-known HOM dip \cite{hom1987OG,ou1988hom}. We mathematically describe this contrivance as follows. %We present the coincidence probability p_c for independent Gaussian wavepackets as inputs to the HOM device, leaving explicit calculations to the Appendix... 

Consider a two-photon source, which generates independent single photons, in spatial modes $(a,b)$ and occupying wavepackets $(f,g)$, such that initial two-photon state is 
\begin{equation}
\begin{split}
    \ket{\Psi}&=\hat{a}_f^\dagger\hat{b}_g^\dagger\ket{\mathbf{0}} \\ 
    &=\left(\int\int\dd\nu_1\dd\nu_2 \ f_1g_2\hat{a}_1^\dagger\hat{b}_2^\dagger\right)\ket{\mathbf{0}},
    \end{split}
\end{equation}
where subscripts indicate spectral dependence. Introducing a relative time delay $\delta t$ in, say, mode $a$, followed by the 50/50 beamsplitter transformation, 
\begin{align*}
    \hat{a}^\dagger&\longrightarrow \frac{1}{\sqrt{2}}\left(\hat{a}^\dagger + \hat{b}^\dagger\right) \\ 
     \hat{b}^\dagger&\longrightarrow \frac{1}{\sqrt{2}}\left(\hat{b}^\dagger - \hat{a}^\dagger\right),
\end{align*}
leads to the state
\begin{equation}
\begin{split}
    \ket{\Psi_{\delta t}}=\frac{1}{2}\int & \dd\nu_1\dd\nu_2 \ f_1g_2\text{e}^{\text{i}\nu_1\delta t} \\ &\times\left[\left(\hat{a}_1^\dagger\hat{b}_2^\dagger-\hat{a}_2^\dagger\hat{b}_1^\dagger\right)-\left(\hat{a}_1^\dagger\hat{a}_2^\dagger -\hat{b}_1^\dagger\hat{b}_2^\dagger\right)\right]\ket{\mathbf{0}},
\end{split}
\end{equation}
with the first parenthetic term causing coincident detection events and the second term embodying photon bunching. Note that, for identical monochromatic wavepackets (e.g., equivalent Dirac-delta distributions for the spectral functions), the first term vanishes -- independent of any delay -- and both photons come out in mode $a$ or mode $b$. 

To calculate the likelihood of single-photon coincidence events, we introduce the single-photon projection operators
\begin{align}
    \hat{\Pi}_a&=\int\dd\nu \ \hat{a}_\nu^\dagger\dyad{\mathbf{0}}\hat{a}_\nu \\ 
    \hat{\Pi}_b&=\int\dd\nu \ \hat{b}_\nu^\dagger\dyad{\mathbf{0}}\hat{b}_\nu. 
\end{align}
The coincidence detection probability $p_c$, as a function of the time delay, is then given by 
\begin{equation}
    p_c(\delta t) = \ev{\hat{\Pi}_a\otimes\hat{\Pi}_b}{\Psi_{\delta t}}.
\end{equation}
For identical, Gaussian wave-packets ($f_k=g_k$) with spectral width $\sigma$, the coincident probability\footnote{See also refs. \cite{ou2007book,branczyk2017HOMnotes} for meticulous calculations of this independent wave-packet scenario and many more.} reduces to 
\begin{equation}
    p_c(\delta t)=\frac{1}{2}\left[1- \exp(-\frac{1}{2}{\delta t}^2\sigma^2)\right].\label{eq-pc_ideal}
\end{equation}
For $\delta t = 0$, the coincident probability vanishes, since the single-photons jointly exit from one port of the 50/50 beamsplitter or the other, however this is the ideal case -- without noise and with unit visibility.

\subsubsection{Digression on noise: visibility, fluctuations, and gravitational decoherence}\label{subsubsec-noise}

The expressions in the previous section dealt with ideal HOM interference, where one has perfect control over all distinguishability parameters. In reality, however, such is not the case, and one is led to the more general expression for the coincidence probability 
\begin{equation}
    p_c(\delta t)=\frac{1}{2}\left[1- \mathcal{V}\exp(-\frac{1}{2}{\delta t}^2\sigma^2)\right], 
\end{equation}
where $\mathcal{V}$ is the visibility of the interference experiment, defined by 
\begin{equation}
    \mathcal{V}\coloneqq\frac{p^{\text{max}}_c-p^{\text{min}}_c}{p^{\text{min}}_c},
\end{equation}
with
\begin{align*}
    p^{\text{max}}_c&\coloneqq\lim_{\delta t\rightarrow\infty}p_c(\delta t) \\
     p^{\text{min}}_c&\coloneqq\lim_{\delta t\rightarrow0}p_c(\delta t).
\end{align*}
Systematic control allows one to maintain the visibility criterion $\mathcal{V}>1/2$, which demarcates the boundary between sufficient quantum and/or classical descriptions for the fields \cite{ou1988hom,ou2007book}. 

One source of noise -- pertinent to this work -- is temporal fluctuations arising from, say, random changes in the relative paths taken by the photons. One can model this as a background Gaussian-noise, where the time delay is a random variable with mean $\delta t$ and fluctuation $\widetilde{\sigma}$. The mathematical particulars are inconsequential (see, e.g., \cite{demkowicz2015opticalint} for various noise models in optical interferometry); however, since we concern ourselves with gravitationally induced temporal \emph{shifts} in the HOM dip, we require the size of the fluctuations $\widetilde{\sigma}$ to be smaller than the average size of the shift, in order for the gravitational phenomena to be resolvable in practice. Such is easier for, e.g., common-path (Section \ref{subsec-frame_drag_hom}) versus dual-arm (Section \ref{subsec-mzi_hom}) interferometry. 

A more fascinating noise source is that of gravitational/space-time fluctuations, which could, in principle, measurably affect the visibility of optical interference signatures. General relativity treats space-time as a dynamical variable, hence, gravitational fluctuations -- either of classical or quantum origin -- lead to a loss of coherence in quantum systems.\footnote{For all details omitted here and for a more thorough discussion, see ref. \cite{bassi2017decoherence}.} Unlike other sources of noise, which can be eliminated by lowering temperatures and creating extreme vacua, it is impossible to get rid of gravitational decoherence. 

On the classical side, space-time fluctuations can have well-determined and various origins -- from the seismic activity of nearby masses, to the more exotic incoherent superpositions of gravitational waves and/or artifacts of other gravitational sources scattered about the universe. One can model such fluctuations as stochastic waves, with unequal time correlation functions of metric components, which one characterizes with spectral distributions. Then, for example, the geometry of an interferometer along with these spectral distributions determine the visibility of an interference signature and can thus, in principle, have observable manifestations \cite{bassi2019decoherence}.

On the quantum side, however, a fully determined and consistent understanding of the origins of quantum space-time fluctuations is far from satisfactory. Nevertheless, several phenomenological models have been proposed, including treatments where one links the collapse of the wave-function to gravitational decoherence or semi-classical treatments, which replace the sources in Einstein's field equations with the expectation value of the stress-energy tensor operator. These models generally rely on parameters determining the scale at which quantum-gravity effects become relevant, i.e. the Planck scale, however this scale is discouragingly minuscule (Planck length, $\ell_P\sim10^{-35}\text{m}$). Therefore, it is difficult to imagine that quantum-gravitational fluctuations will become prevalent in interferometry scenarios, in the near to mid-term. Though, with the continuing development of large-scale interferometers and ever-improving sensitivities, perhaps bounds may be put on these scales, in the future.

\section{Setup and Results}\label{subsec-setup_results}

\subsection{Relativistic frame-dragging and the HOM dip}\label{subsec-frame_drag_hom}

\begin{figure}
    \centering
    \includegraphics[width=\linewidth]{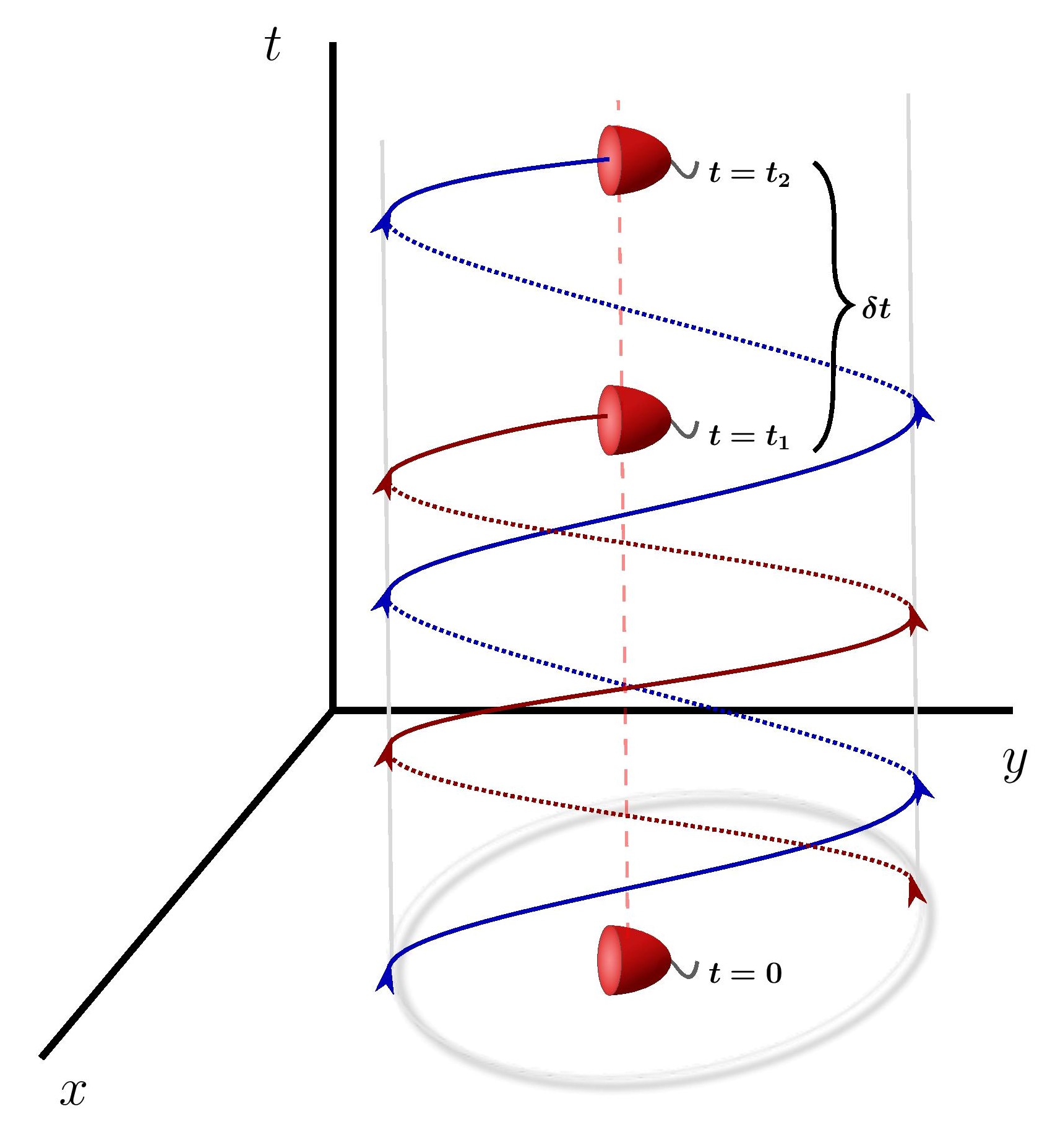}
    \caption{Space-time diagram of counter-propagating null rays. Two electromagnetic signals (red and blue curves) counter-propagate along a common path in space. Relativistic frame-dragging effects cause the pitch of the space-time helices to differ, which leads to differing arrival times at a stationary detector system. All motion is constrained to a world-tube of spatial size $\ll\ell\sim \min(c^2/g,c/\omega^\prime)$.}
    \label{fig-spacetime_path}
\end{figure}

Consider a HOM-interference experiment, wherein two single-photon wavepackets counter-propagate through a common-path interferometer constrained to the earth's surface (FIG. \ref{fig-earth_hom}). Relativistic effects -- e.g., the Sagnac, Lense-Thirring, and geodetic effects -- induce distinguishability between the counter-propagating paths, leading to a temporal shift in the HOM dip, which would otherwise be at $\delta t =0$ if relativistic effects were absent.

The off-diagonal terms of the metric perturbation -- physically corresponding to rotational motion -- are the sole contributors to the time delay, \textit{ceteris paribus}, which one can calculate via concurrent use of equations \eqref{eq-phase_soltn} and \eqref{eq-prop_metric}, in the proper reference frame of an observer. For a single path around the interferometer loop -- say, the right-handed path -- one finds

\begin{equation}
    \begin{split}
    \delta t_{\text{RH}} = c^{-1}\ointctrclockwise \dd x^{(i)} h_{(0)(i)}&= c^{-2}\ointctrclockwise \dd \vec{x}\vdot(\vec{\omega}^\prime\cross\vec{x}) \\
    &=c^{-2}\int\dd\vec{\mathcal{A}}\vdot\grad\cross(\vec{\omega}^\prime\cross\vec{x}) \\
    &=\frac{2\vec{\omega}^\prime\vdot\vec{\mathcal{A}}}{c^2},
    \end{split}
\end{equation}
where $\vec{\mathcal{A}}$ is the areal vector, perpendicular to the interferometry plane, and we have used the spatial-coordinate independence of $\omega^\prime$. A similar relation holds for the left-handed path, albeit with opposite sign, such that
\begin{equation}
    \delta t\coloneqq\delta t_{\text{RH}}-\delta t_{\text{LH}} =\frac{4\vec{\omega}^\prime\vdot\vec{\mathcal{A}}}{c^2},
\end{equation}
with $\vec{\omega}^\prime$ given by equation \eqref{eq-rotation_rate}. 

\begin{figure*}[t]
    \includegraphics[width=1\textwidth]{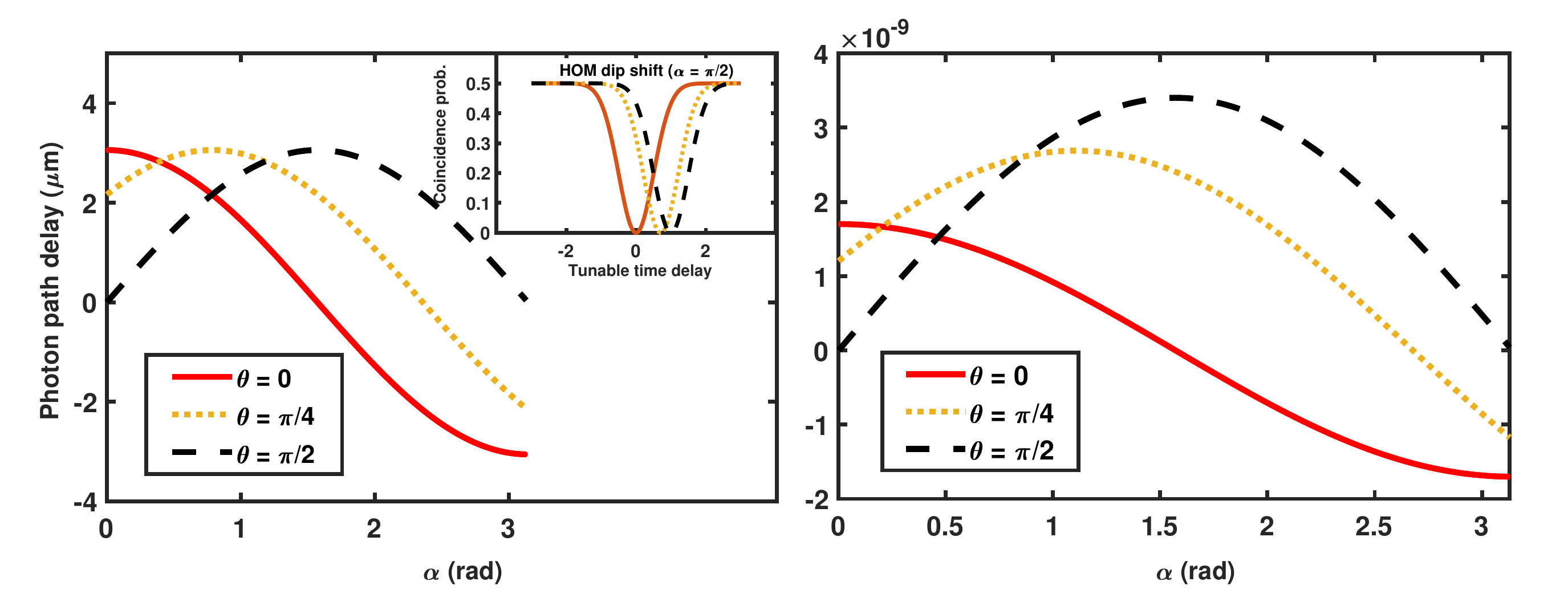}
    \caption{Photon path delay ($c\delta t$, where $c$ is the speed of light) as a function of the interferometer-orientation angle $\alpha$, for a common-path configuration, with $\mathcal{A}=1\text{km}^2$. Solid (red), dotted (yellow), and dashed (black) lines represent various latitudes $\theta$ at which a detector is placed. \textbf{Left:} Path delay due to the Sagnac effect. The inset is a representative figure for the HOM-dip shift (the spectral widths have been exaggerated for visual clarity) due to the relative delay between the counter-propagating modes, at a fixed orientation angle $\alpha=\pi/2$ and at various latitudes $\theta$. Similar results hold for, e.g., a fixed latitude and various orientation angles. This latter scenario being more practical, as one would be at a fixed position on earth ($\theta=\text{constant}$) whilst performing the experiment at several interferometer-orientation settings (various $\alpha$).  \textbf{Right:} Path delay due to the combined Lense-Thirring and geodetic effects.}
    \label{fig-sagnac_results}
    \end{figure*}

To evaluate the above expression, for an observer constrained to the earth's surface, we choose an earth-centered spherical-coordinate system $(\hat{\e}_r,\hat{\e}_\theta,\hat{\e}_\phi)$ co-moving with the observer, with $\vec{\omega}$ aligned along the polar axis.\footnote{Thus, $\vec{v}=\vec{\omega}\cross\vec{R}$ is the observer's coordinate velocity} The earth's radius $R$ and the observer's latitude $\theta$ establishes the earthbound constraint, since -- due to the intrinsic axisymmetric structure of the background space-time -- characterization of the observer's longitude is superfluous. Furthermore, we assume the areal vector to take the form $\vec{\mathcal{A}}=\mathcal{A}(\hat{\e}_r\cos{\alpha}-\hat{\e}_\theta\sin{\alpha})$, with $\alpha$ being the angle between the observer's normal and the areal vector of the interferometer. Under these considerations, the coordinate expression for the time delay is then 
\begin{widetext}
\begin{equation}
    \delta t = \underbrace{\frac{4\omega \mathcal{A}}{c^2}\cos(\theta - \alpha)}_{\text{Sagnac}} +\underbrace{\frac{4\omega \mathcal{A}}{c^2}\left(\frac{2GM}{c^2R}\sin\theta\sin\alpha+\frac{GI}{c^2R^3}(2\cos\theta\cos\alpha-\sin\theta\sin\alpha)\right)}_{\text{geodetic + Lense-Thirring}},
\end{equation}
\end{widetext}
which agrees with \cite{gingerPRD}. We note that the general relativistic contributions are on the order of $r_S/R\sim10^{-9}$ times smaller than the Sagnac contribution, where $r_S=2GM/c^2$ is the Schwarzschild radius of the earth.

We now substitute the above expression into the coincidence probability formula \eqref{eq-pc_ideal} and plot the results for various parameter values $(\theta,\mathcal{A},\alpha)$. See FIG. \ref{fig-sagnac_results} for analysis.   

\subsubsection{Order-of-magnitude estimates}\label{subsec-estimates_1}

Perhaps more insightful than a full-blown analysis is an order-of-magnitude estimate for these effects. A key quantity to then consider is the size of the temporal shift relative to the physical size of the interferometer; thus, define \begin{equation}
\mathcal{F}\coloneqq\frac{\delta t^\star}{\mathcal{A}},\label{eq-eff}
\end{equation}
which has dimensions s/km$^2$ and where $\delta t^\star$ is crudely the difference between the maximum and minimum time delay. This quantity is of practical interest as e.g., given some physical constraint on the areal dimensions, one can see what level of precision is required in order to observe relativistic effects.

As example, the quantity $\mathcal{F}$, for the Sagnac effect, goes as \[\mathcal{F}_\text{Sag}\sim\frac{4\omega}{c^2}\sim10^{-15} \ \text{s}/\text{km}^{2},\] and since the Lense-Thirring and geodetic effects are a billionth of the size of the Sagnac effect, we have\[\mathcal{F}_\text{GR}\sim10^{-9}\mathcal{F}_\text{Sag}\sim10^{-24} \ \text{s}/\text{km}^{2},\] where the subscript GR concurrently signifies the Lense-Thirring and geodetic effects. 

The smallness of the Lense-Thirring and geodetic effects is immediately concerning, as it implies stringent experimental constraints (high-precision and control of a large-area interferometer), however, this is not to say that observation of such is impractical. On the contrary, there is ongoing research towards terrestrially measuring these frame-dragging effects with \emph{classical}-optical interference \cite{gingerPRD,ginger2020}. With regards to a similar undertaking in HOM interferometry, a full feasibility analysis is duly wanting.\footnote{Perhaps at the level of \cite{hilweg2017}. One should also leverage parameter estimation techniques to optimize operating conditions (see ref. \cite{lyons2018attosecond}), but such is beyond the scope of this work.}      

On the other hand, observing signatures of the Sagnac effect (induced by the earth's rotation) via HOM interference appears approachable. Recently, phenomena akin to this were observed by Restuccia et al. \cite{padgett2019hom} (and analyzed further in \cite{padgett2020nir}), wherein the authors constructed a HOM configuration (FIG. \ref{fig-earth_hom}) upon a rotating table with tunable rotation rate, and subsequently discovered a rotation-induced shift in the HOM dip (proportional to the rotation frequency), as one would predict with the formalism presented here. To measure an analogous effect induced by the earth's rotation, one requires much greater time-delay precision, due to the minute rotational frequency of the earth $\omega\sim10^{-5}\text{s}^{-1}$, which seems presently achievable with modest resources. For example, considering a fiber-loop of radius $r$ and $N$ turns, the effective area of the interferometer is $\mathcal{A}=N\pi r^2$ or $\mathcal{A}=lr/2$, where $l=2\pi Nr$ is the length of the fiber. Taking $l\approx2$km and $r\approx1\text{m}$ implies $\mathcal{A}_\text{Sag}\sim10^{-3}\text{km}^2$, which in turn implies a required time-delay precision at the attosecond ($1\text{as}=10^{-18}\text{s}$) scale -- on a par with current experiment \cite{lyons2018attosecond}. Proof-of-principle demonstrations like these are quite intriguing, as they constitute quantum fields in non-inertial reference frames.

\subsection{Gravitational time-dilation and the HOM dip}\label{subsec-mzi_hom}

\begin{figure}[]
    \centering
    \includegraphics[width=\linewidth]{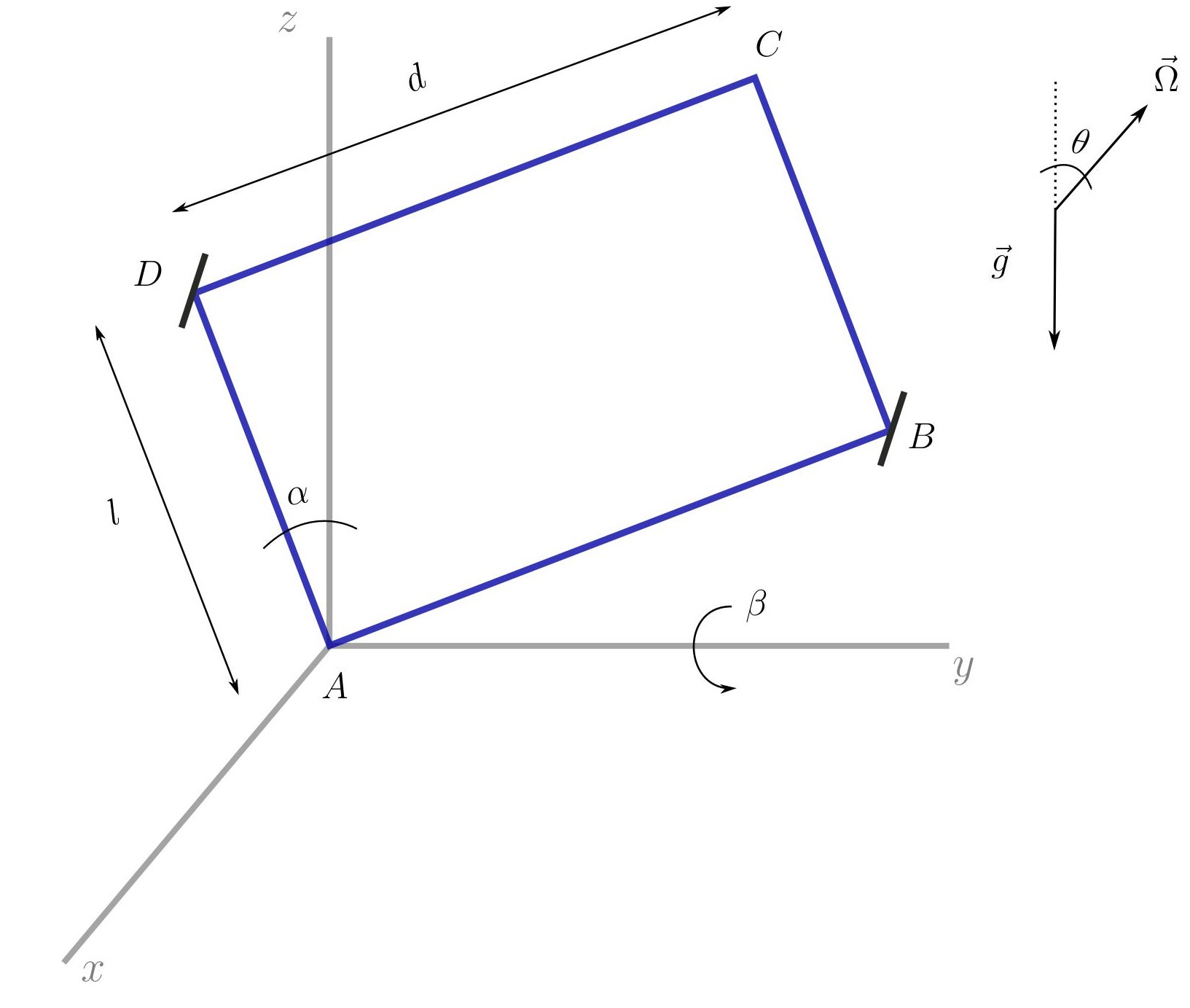}
    \caption{A dual-arm configuration with a two-photon source (or, a beamsplitter and two photon-detectors) at $C$, mirrors at $B/D$, and a beamsplitter and two photon-detectors (or, a two-photon source) at the origin $A$. The local reference frame is fixed to the earth's surface at a latitude $\theta$. The $z$-axis is parallel to the radial vector $\hat{\e}_r$, and the $y$-axis points along the longitudinal line $\hat{\e}_\phi$ at the observer's position. The locally measured angles, $\alpha$ and $\beta$, determine the orientation of the interferometer, with respect to the horizontal and vertical planes. The area of the interferomter is $\mathcal{A}=ld$.}
    \label{fig-mzi_hom}
\end{figure}

We now consider a HOM-interference experiment, for a dual arm configuration, and calculate the time delay due to the effects of the uniform gravitational acceleration and centrifugal acceleration measured by an observer fixed upon earth's surface (see FIG. \ref{fig-mzi_hom}). We ignore the off-diagonal rotation-induced contributions, as they are equivalent (up to a factor of 2) to the calculations of the previous section. 

The unit vectors, along each arm of the interferometer, are given by
\begin{align*}
\hat{n}_{BA} &= \hat{n}_{CD} = (\sin{\alpha}\sin{\beta}, \cos{\alpha}, \sin{\alpha}\cos{\beta}) \\
\hat{n}_{DA} &= \hat{n}_{CB} = (\cos{\alpha}\sin{\beta}, -\sin{\alpha}, \cos{\alpha}\cos{\beta}).
\end{align*}
The coordinate equations for the lines along the arms are thus,
\begin{align*}
BA &\qq*{:}\frac{x}{d\sin{\alpha}\sin{\beta}} = \frac{y}{d\cos{\alpha}} = \frac{z}{d\sin{\alpha}\cos{\beta}}\\
CD &\qq*{:} \frac{x - l\cos{\alpha}\sin{\beta}}{d\sin{\alpha}\sin{\beta}} = \frac{y + l\sin{\alpha}}{d\cos{\alpha}} = \frac{z - l\cos{\alpha}\cos{\beta}}{d\sin{\alpha}\cos{\beta}} \\
DA &\qq*{:} \frac{x}{l\cos{\alpha}\sin{\beta}} = \frac{y}{-l\sin{\alpha}} = \frac{z}{l\cos{\alpha}\cos{\beta}} \\
CB &\qq*{:}\frac{x - d\sin{\alpha}\sin{\beta}}{l\cos{\alpha}\sin{\beta}} = \frac{y - d\cos{\alpha}}{-l\sin{\alpha}} = \frac{z-d\sin{\alpha}\cos{\beta}}{l\cos{\alpha}\cos{\beta}}
\end{align*}
By concurrent use of equations \eqref{eq-phase_soltn} and \eqref{eq-prop_metric}, we calculate the phase difference, $\delta S$, accrued along a path due to the uniform acceleration $\vec{\gamma}$,
\begin{equation}
\begin{split}
\delta S &= -\frac{1}{2} \int \dd x^{(0)}k^{(0)}h_{(0)(0)} \\
&=\int\left(\dd\vec{x}\vdot\vec{k}\right)\vec{\gamma}\vdot\vec{x},
\end{split}\label{eq-phase_accel}
\end{equation}
For example, setting $\vec{\gamma} = g\hat{\e}_z$, where $g\approx9.8\text{m}/\text{s}^2$ is the gravitational acceleration on the earth's surface, we find the phase along the segment $BA$ to be $$\delta S_{BA}  = gk^{(0)} \int^A_B (\vec{dx}\vdot \hat{n}_{BA}) z,$$ where $\hat{n}_{BA}$ is the unit vector along $BA$. Integrating along the each arm of the interferometer, we then obtain
\begin{align*}
\delta S_{BA} &= k^{(0)} \frac{g}{2c^3}d^2 \sin{\alpha}\cos{\beta}\\
\delta S_{CB} &=  k^{(0)} \frac{g}{c^3}\left(\frac{l^2}{2} \cos{\alpha}\cos{\beta} + ld \sin{\alpha}\cos{\beta}\right)\\
\delta S_{CD} &= k^{(0)} \frac{g}{c^3}\left(\frac{d^2}{2} \sin{\alpha}\cos{\beta} + ld \cos{\alpha}\cos{\beta}\right)\\
\delta S_{DA} &= k^{(0)} \frac{g}{2c^3}l^2 \cos{\alpha}\cos{\beta}.
\end{align*}

Since the observer is fixed to the earth's surface (a non-inertial reference frame), they also measure a centrifugal acceleration 
\[\vec{\mathfrak{a}}=\omega^2 R\sin{\theta} (\sin{\theta} \hat{\e}_z + \cos{\theta} \hat{\e}_y).\] Then, setting $\vec{\gamma}=\vec{\mathfrak{a}}$ in equation \eqref{eq-phase_accel}, we find additional phase differences along the different interferometer arms,
\begin{align*}
\delta S_{BA} &=  k^{(0)} \frac{\omega^2 R }{2c^3}d^2\sin{\theta} (\cos{\theta}\cos{\alpha} + \sin{\theta}\sin{\alpha}\cos{\beta})\\
\delta S_{CB} &=  k^{(0)} \frac{\omega^2 R}{c^3} \sin{\theta} \Bigg[\sin{\theta}\cos{\beta}\left(\frac{l^2}{2}\cos{\alpha} + ld\sin{\alpha}\right) \\ &\qq{}\qq{}\qq{}  \qq{} \ -\cos{\theta}\left(\frac{l^2}{2}\sin{\alpha} + ld\cos{\alpha}\right)\Bigg]\\
\delta S_{DC} &=  k^{(0)} \frac{\omega^2 R}{c^3} \sin{\theta} \Bigg[\cos{\theta}\left(\frac{d^2}{2}\cos{\alpha} - ld\sin{\alpha}\right)\\ & \qq{}\qq{}\qq{}  + \sin{\theta}\cos{\beta}\left(\frac{d^2}{2}\sin{\alpha} + ld\cos{\alpha}\right)\Bigg]\\
\delta S_{DA} &=  k^{(0)} \frac{\omega^2 R }{2c^3}l^2\sin{\theta}\left( \sin{\theta}\cos{\alpha}\cos{\beta} - \cos{\theta}\sin{\alpha}\right).
\end{align*}
Therefore, considering identical single-photons traversing separate paths $ABC$ and $CDA$, we obtain the total time delay between the paths (after dividing by the mean-frequency $k^{(0)}$), due to both the gravitational and centrifugal accelerations,
\begin{widetext}
\begin{equation}
    \delta t = \underbrace{\frac{g\mathcal{A}}{c^3} \cos{\beta}(\cos{\alpha}-\sin{\alpha})}_{\text{gravitational, $g$}} +\underbrace{\frac{\omega^2 R\mathcal{A}}{c^3}\sin\theta\left[\sin{\theta} \cos{\beta}(\cos{\alpha}-\sin{\alpha}) + \cos{\theta} (\cos{\alpha}+\sin{\alpha})\right]}_{\text{centrifugal, $\mathfrak{a}$}},
\end{equation}
\end{widetext}
where $\mathcal{A}=ld$ is the interferometer area. We note that $\mathfrak{a}/g\sim10^{-2}$. Plots for these results are shown in FIG. \ref{fig-mzi_hom}.

\begin{figure*}[t]
    \includegraphics[width=1\textwidth]{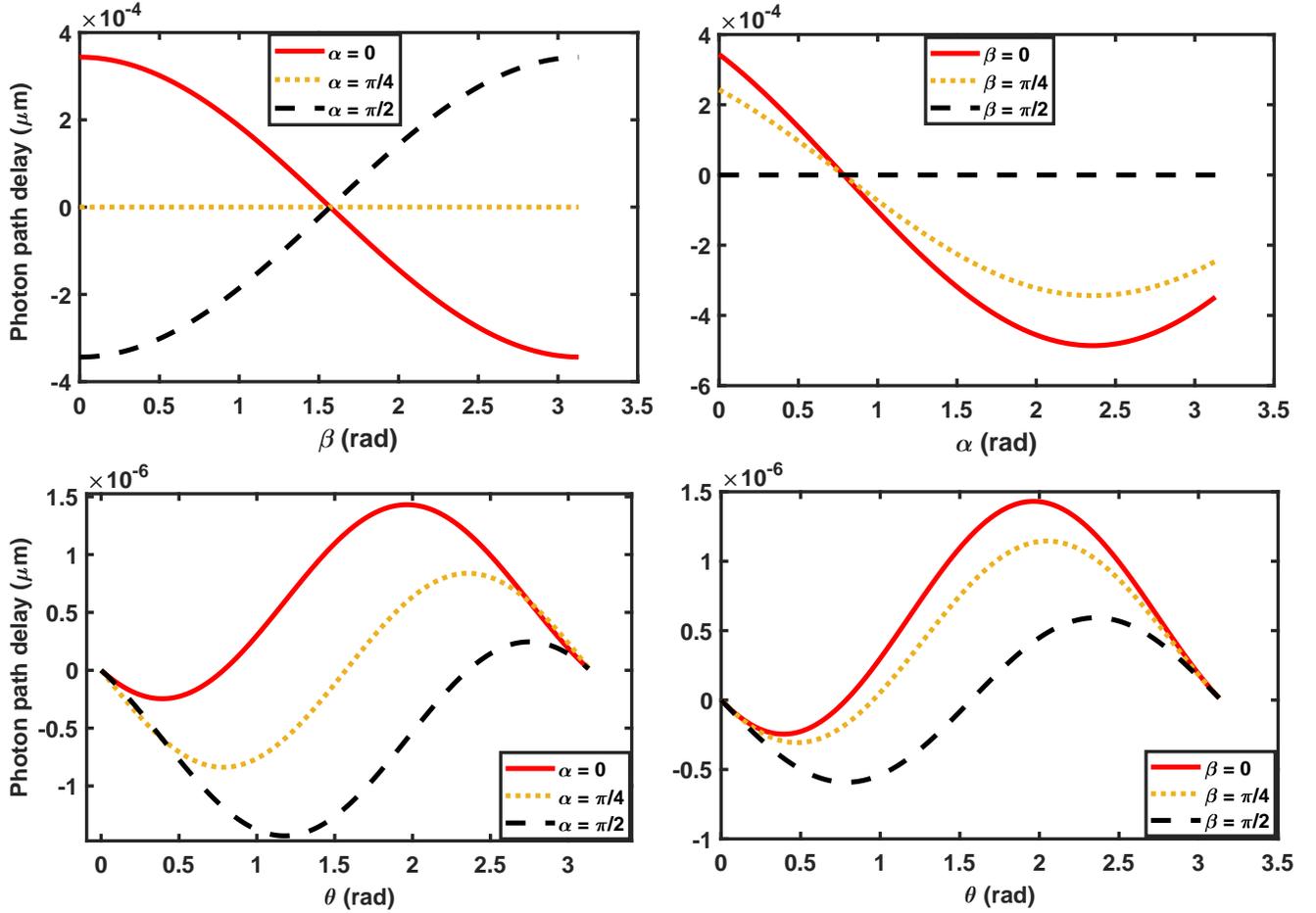}
    \caption{Photon path delay ($c\delta t$), for a dual-arm configuration, with $\mathcal{A}=1\text{km}^2$. \textbf{Top left (right):} Path delay induced by the gravitational acceleration $g\approx9.8\text{m}/\text{s}^2$, as a function of the orientation angle $\beta$ ($\alpha$). \textbf{Bottom left (right):} Path delay induced by the centrifugal acceleration $\mathfrak{a}\approx3.4\times10^{-2}\text{m}/\text{s}^2$, as a function of the latitude $\theta$, for various orientation angles $\alpha$ ($\beta$) and with $\beta=0$ ($\alpha=0$) for all curves.}
    \label{fig-mzi_theta}
    \end{figure*}

\subsubsection{Order-of-magnitude estimates}\label{subsec-estimates_2}
Consider the photon time delay per unit area, characterized by the quantity $\mathcal{F}$ [eq. \eqref{eq-eff}] and induced by gravitational and centrifugal accelerations. For the gravitational acceleration, we have \[\mathcal{F}_g\sim\frac{g}{c^3}\sim10^{-19} \ \text{s}/\text{km}^{2},\] which is $10^3$ times smaller than the Sagnac effect and $10^6$ times larger than the minute geodetic and Lense-Thirring effects considered in Section \ref{subsec-estimates_1}. Similarly, for the centrifugal acceleration, we have\[\mathcal{F}_\mathfrak{a}\sim\frac{\omega^2 R}{c^3}\sim10^{-21} \ \text{s}/\text{km}^{2}.\]

Note that Lyons et al. \cite{lyons2018attosecond} recently acheived HOM time-delay resolution at the attosecond scale (1as=$10^{-18}$s) -- closely approaching the scale implied by $\mathcal{F}_g$ -- but with a much smaller interferometer area than required here. Nevertheless, a tabletop HOM experiment of this sort seems achievable with present or near-term technology. Indeed, Hilweg et al. \cite{hilweg2017} investigated the feasibility of a similar experiment aimed at measuring gravitational time-dilation effects via single-photon interference, with a variant Mach-Zehnder interferometer, and found that, though challenging, a tabletop experiment is possible with long-fiber spools and active phase stabilisation. An analogous feasbility analysis for a HOM experiment -- leveraging parameter estimation techniques, like those used in \cite{lyons2018attosecond} -- should be pursued. We leave this for future work.  

\section{Summary and Conclusion}\label{sec-conclusion}
In this work, we showed how to encode general-relativistic effects -- e.g., frame-dragging and gravitational time-dilation effects -- into multi-photon quantum-interference phenomena, for various interferometer configurations. This was theoretically achieved by quantizing a massless scalar-field in a weak gravitational field (geometric quantum-optics in curved space-time, in the weak field regime; Section \ref{sec-qu_optics}). Applying this formalism to a terrestrial laboratory setting (Section \ref{subsec-metric}), we showed that, in principle, gravitational effects can induce observable changes in quantum-interference signatures, using HOM interference as an exemplar (Sections \ref{subsec-hom} and \ref{subsec-setup_results}). Non-inertial effects, due to the earth's rotation and centrifugal acceleration, were also considered, and the potentiality of practical demonstrations was briefly analyzed. Though the latter analysis was unfulfilling (a full feasibility, in this regard, is still wanting), the landscape of potential proof-of-principle demonstrations, like these, appears promising -- the reason being that analogous enterprises exploring, e.g., gravitational time-dilation via single-photon interference, with tabletop long-fiber spools \cite{hilweg2017} or with quantum satellites \cite{terno2018proposal}, have been pursued and found feasible with current technology. Further inspiration for this potentiality comes from concurrent consideration of recent experimental endeavors: a HOM-interference experiment in a rotating, non-inertial reference frame \cite{padgett2019hom}; high-precision HOM time-delay resolution \cite{lyons2018attosecond}; and current attempts to measure relativistic frame-dragging effects with a classical, optical, earthbound system \cite{gingerPRD,ginger2020}. 

From historical considerations, one may view our result as a quantum analog to the seminal proposal of Scully et al. \cite{scully1981} -- made nearly forty years ago -- who originally investigated the potentiality of observing PN (e.g., general-relativistic frame-dragging) effects with \emph{classical} optical interferometry. In a similar fashion, a proof-of-principle demonstration of our results would serve as a genuine \emph{quantum} PN test of general relativity -- extending the domain of validity of Einstein's classical theory of gravitation, to the arena of quantized electromagnetic fields.

\acknowledgements

AJB and SH acknowledge Jonathan P. Dowling for many fruitful conversations and his ever-endearing support. AJB acknowledges financial support from the NSF.   

\bibliography{Photon_bunching_and_Gravitomagnetism}
\end{document}